\documentclass[twocolumn, superscriptaddress, amsfonts, floatfix]{revtex4} 

\usepackage{graphicx}
\usepackage{dcolumn}
\usepackage{bm}
\usepackage{color}

\begin{document}

\preprint{APS/123-QED}

\title{Field-induced anomaly in the anisotropic non-Fermi-liquid normal state  of UBe$_{13}$
}

\author{Yusei Shimizu}
\email{yshimizu@issp.u-tokyo.ac.jp}
\affiliation{Institute for Solid State Physics (ISSP), University of Tokyo, Kashiwa, Chiba 277-8581, Japan}
\author{Shunichiro Kittaka}
\affiliation{Department of Basic Science, The University of Tokyo, Meguro, Tokyo 153-8902, Japan}
\author{Yohei Kono}
\affiliation{Department of Physics, Chuo University, Kasuga, Bunkyo-ku, Tokyo, 112-8551, Japan}
\author{Shota Nakamura}
\affiliation{Nagoya Institute of Technology, Aichi, Nagoya 466-8555, Japan}
\author{Yoshinori Haga}
\affiliation{Advanced Science Research Center, Japan Atomic Energy Agency, Tokai, Ibaraki,  319-1195,  Japan. }
\author{Etsuji Yamamoto}
\affiliation{Advanced Science Research Center, Japan Atomic Energy Agency, Tokai, Ibaraki,  319-1195,  Japan. }
\author{Kazushige Machida}
\affiliation{Department of Physics, Ritsumeikan University, Kusatsu, Shiga 525-8577, Japan.}
\author{Hiroshi Amitsuka} 
\affiliation{Graduate School of Science, Hokkaido University, Sapporo, Hokkaido,  060-0810, Japan.}
\author{Toshiro Sakakibara} 
\affiliation{Institute for Solid State Physics (ISSP), University of Tokyo, Kashiwa, Chiba 277-8581, Japan}

\date{\today}%

\begin{abstract}
We report the results of high-resolution  dc magnetization and specific-heat measurements at very low temperatures for
 a single crystal \color{black}  of UBe$_{13}$ in magnetic fields applied along the [001] and [111] directions, in both the normal and superconducting states. In the normal state, magnetic susceptibility $\chi(T) = M/H$ exhibits a logarithmic temperature dependence over a wide temperature range (1--20 K). However, with increasing  field, this non-Fermi-liquid (NFL) behavior of $\chi(T) $ at low temperatures is suppressed. Moreover, a susceptibility maximum occurs below 4 T, whereas Fermi-liquid coherence is recovered above 8 T. In addition,  thermodynamic anomalies ($T_{\rm A}$ and $H_{\rm A}$) occur in both magnetic susceptibility and specific heat at intermediate fields (6--10 T) along the [111] direction. Furthermore, a nontrivial fifth-order nonlinear susceptibility is observed in the normal-state magnetization of UBe$_{13}$.  These results suggest  a close  relationship between the field-induced multipolar correlations of $5f$-electron degrees of freedom and the Fermi-surface reconstruction accompanying the crossover from the NFL state to the Fermi-liquid state in UBe$_{13}$.
\color{black}
\end{abstract}

\maketitle

\section{Introduction}

Discovered over 40 years ago, UBe$_{13}$ is the first uranium-based heavy-electron superconductor \cite{Ott_PRL_1983}. However, its superconducting and anomalous normal states remain enigmatic \cite{Stewart_JLTP_2019}. As UBe$_{13}$ is a candidate spin-triplet superconductor, extensive studies have been conducted to clarify its superconducting gap symmetry \cite{Ott_PRL_1984, MacLaughlin_PRL_1984, Einzel_PRL_1986, Tien_PRB_1989, Golding_PRL_1985} and the behavior of its upper critical field $H_{\rm c2}$ \cite{Maple_PRL_1985, Fomin_JLTP_2000, Shimizu_PRL_2019}. Unexpectedly, angle-resolved specific-heat measurements have revealed the absence of nodal quasiparticle excitations in UBe$_{13}$ \cite{Shimizu_PRL_2015}. These observations have prompted reconsideration of the pairing mechanism in heavy-electron systems, where nodal gap symmetries are expected. Moreover, UBe$_{13}$ exhibits a magnetic anomaly ($B^{*}$ anomaly) in its superconducting state \cite{Ellman_PRBR_1991, Kromer_PRL_1998, Langhammer_JMMM_1998, Walti_PRBR_2001, Kromer_PRB_2000, Shimizu_PRL_2012, Shimizu_PRB_2016}. Although the origin of this anomaly remains unresolved, it is deemed a precursor to the double transition in the multiple superconducting phases of U$_{1-x}$Th$_{x}$Be$_{13}$ (0.019 $\lesssim x \lesssim$ 0.045) \cite{Ott_PRBR_1985, Ott_PRB_1986, Kromer_PRL_1998, Kromer_PRB_2000}.

This study focuses on the anomalous normal state of UBe$_{13}$. This material shows non-Fermi-liquid (NFL) behavior in various physical properties, including electrical resistivity \cite{Brison_JdePhysique_1989}, specific heat \cite{Mayer_PRB_1986, Gegenwart_PhysicaC_2004}, magnetic susceptibility \cite{McElfresh_PRB_1993}, and thermoelectric power \cite{Shimizu_PRBR_2015}. The origin of this NFL behavior is unclear, and several scenarios have been proposed, including a quadrupolar Kondo effect based on the $\Gamma_{3}$ crystal-electric-field (CEF) ground state in the $5f^2$ (U$^{4+}$, $J =$ 4) configuration \cite{Cox_PRL_1987, Cox_JPhysC_1996}, a magnetic-field-induced antiferromagnetic quantum critical point \cite{Gegenwart_PhysicaC_2004, Schmiedeshoff_PhilosMagazine_2009}, and a theoretical model involving competition between the Kondo--Yosida singlet and the $\Gamma_{1}$ CEF singlet in the $5f^2$ configuration \cite{Nishiyama_JPSJ_2011}. A recent study proposed that the semi-metallic character of the conduction electrons (electron and hole Fermi surfaces) originating from the surrounding beryllium atoms may induce the NFL behavior with the two-channel Kondo effect \cite{Iimura_PRB_2019}.

To gain deeper insights into the NFL behavior in UBe$_{13}$, we performed high-resolution dc magnetization and specific-heat measurements on a single crystal of UBe$_{13}$ at very low temperatures. According to recent reports, when single crystals are prepared using the Al-flux method, the Al inclusions in the sample affect the superconducting properties of the crystals \cite{Volts_PhilosophicalMag_2018, Amon_ScientificRep_2018}. However, single crystals are difficult to obtain using any other method, such as the arc-melt method \cite{Shimizu_PRBR_2017}. Therefore, Al-flux-grown single-crystal samples are needed to examine anisotropic effects. Focusing on the anisotropy between $H ||$ [001] and [111] in this study, we examine the precise temperature and field dependence of the magnetization and specific heat of UBe$_{13}$ in its superconducting and normal states. In particular, we report novel field-induced anomalies and remarkable anisotropy in its normal state.

\color{black}

\section{Experimental Procedures}

A single crystal of UBe$_{13}$ was grown via the Al-flux method \cite{Haga_Physica_1999}, and its crystallographic axes were determined using the X-ray Laue method. The sample (6.6 mg) was the same as that used in previous works \cite{Shimizu_PRL_2012, Shimizu_PRB_2016}. Low-temperature magnetization was measured down to 80 mK using homemade capacitive-detection magnetometers \cite{Sakakibara_JJAP_1994, Shimizu_RSI_2021} installed in a $^{3}$He--$^{4}$He dilution and $^{3}$He refrigerators. Magnetic fields up to 14.5 T were applied along the cubic [001] and [111] axes with a field gradient ($G$ = 5 or 8 T/m). For comparison, dc magnetization at 2--370 K was measured using a commercial SQUID magnetometer (MPMS, Quantum Design, Inc.). Low-temperature specific heat was measured using the standard quasi-adiabatic heat-pulse method in the $^{3}$He--$^{4}$He dilution refrigerator down to 80 mK and at magnetic fields of up to 14.5 T applied along the [001], [111], and [110] axes. 
 Here,  magnetic field is presented  in tesla for clarity, whereas the magnetic   susceptibility  is shown  in 
 emu mol$^{-1}$Oe$^{-1}$  for practical convenience and for comparison with previous reports (1 T = 10 kOe) \cite{McElfresh_PRB_1993, Tou_JPSJ_2007}.

\begin{figure}[t]
\begin{centering}
\includegraphics[width=8.8cm]{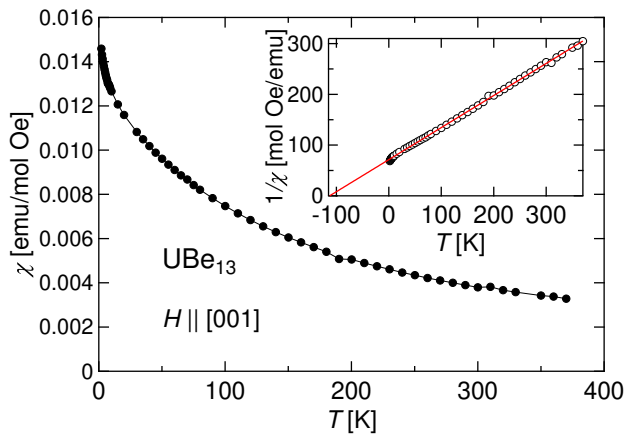}
\caption{   
Temperature dependence of magnetic susceptibility of UBe$_{13}$ in $H ||$[001] at 0.1 T. Inset: temperature dependence of inverse susceptibility, with solid line representing Curie--Weiss fitting.
 }
 \end{centering}
\end{figure}


\section{Results}

\subsection{Magnetic susceptibility and magnetization of  UBe$_{13}$} 

We show the magnetic susceptibility of UBe$_{13}$ measured between 370 and 2 K along $H ||$ [001] at 0.1 T (Fig.1) and then present the low-$T$ results. The overall behavior of susceptibility agrees well with those in previous reports \cite{McElfresh_PRB_1993, Troc_Bulltin_1971, Tou_JPSJ_2007}. As shown in the inset, the high- $T$ region above 120 K follows the Curie--Weiss law, indicating localized $5f$ electrons. A Curie--Weiss fit above 120 K gives an effective magnetic moment of 3.56 $\mu_{\rm B}$/U and a Weiss temperature of $\Theta = -$112 K. This effective moment is close to the expected values of 3.58 and 3.62 $\mu_{\rm B}$/U for $J =$ 4 ($5f^{2}$, U$^{4+}$) and $J =$ 9/2 ($5f^3$, U$^{3+}$), respectively. The valence of U cannot be determined solely from the effective moment. The negative Weiss temperature suggests antiferromagnetic correlations. However, such a large negative Curie--Weiss temperature may arise from the combined effects of complex magnetic interactions and multipolar correlations \cite{McElfresh_PRB_1993}.

Next, we present the low-$T$ magnetization measurement results. Figure 2 shows the magnetization curve of UBe$_{13}$ at the lowest temperature of   80 mK  with the magnetic field  applied  along $H ||$ [111]. Hysteresis is clear below 2 T and around 6--8 T, originating from flux pinning in the superconducting state. The hysteresis around 6--8 T corresponds to the \textit{peak effect}, which often occurs in type II superconductors. However, in UBe$_{13}$, the field where the peak effect disappears ($H_{\rm irr}$) coincides well with the upper critical field $H_{\rm c2}$ determined from specific-heat measurements (i.e., $H_{\rm irr} \simeq H_{\rm c2}$). Equilibrium magnetization ($M_{\rm eq}$) is obtained by averaging the magnetization curves taken in fields with increasing ($M_{\rm inc}$) and decreasing ($M_{\rm dec}$) processes (i.e., $M_{\rm eq} = (M_{\rm inc} + M_{\rm dec})/2$). These results are consistent with previous magnetization data down to 0.14 K along $H ||$ [001] and [110] \cite{Shimizu_PRL_2012, Shimizu_PRB_2016}. The following discussion focuses on normal-state magnetization above $H_{\rm c2}$.

\begin{centering}
\begin{figure}[t]
\includegraphics[width=8.8cm]{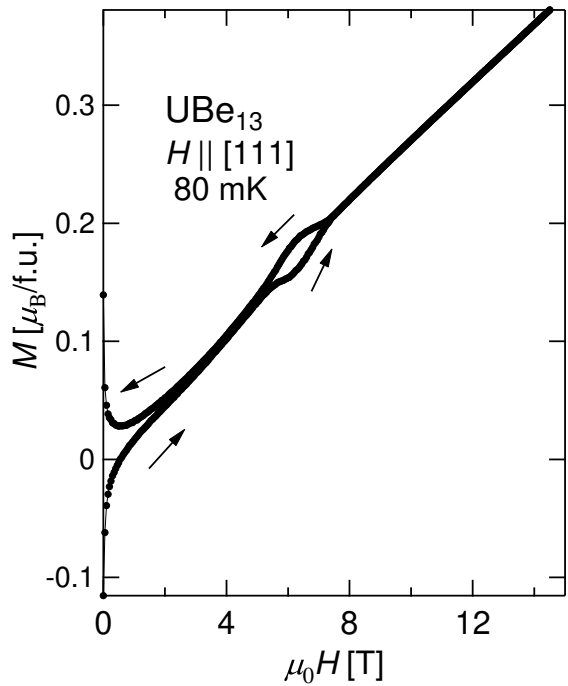}
\caption{   
 Magnetization curve of UBe$_{13}$   at the lowest temperature of 80 mK along $H ||$ [111],
  where the arrows denote the increasing and decreasing processes. 
   }
\end{figure}
\end{centering}

\begin{centering}
\begin{figure}[t]
\includegraphics[width=8.8cm]{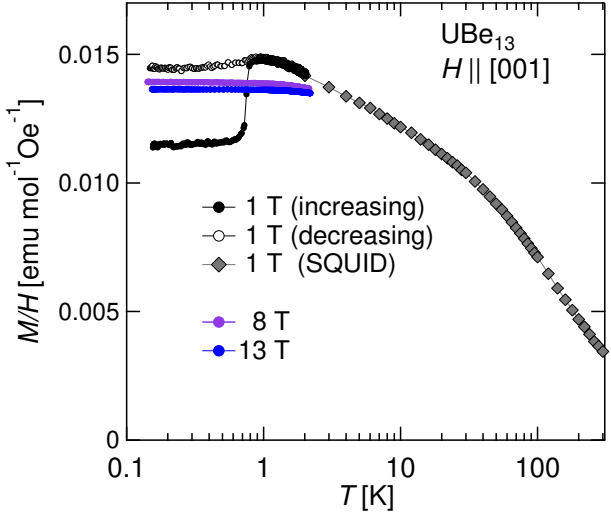}
\caption{   
Temperature dependence of the magnetic susceptibility ($\chi = M/H$) of UBe$_{13}$  for $H ||$ [001] at 1, 8, and 13 T.
 }
\end{figure}
\end{centering}

 Figure 3 shows the temperature dependence of UBe$_{13}$ magnetization along $H ||$ [001] at 1, 8, and 13 T as a function of the temperature logarithm. At 1 T, a superconducting transition occurs in zero-field-cooled (ZFC) and field-cooled (FC) processes. To see the field evolution of the NFL behavior over a wide $T$ range, we also plot the SQUID 
  data measured  between 300 and 2 K at the same field  (1 T, $H ||$ [001]). We observe $\chi \sim -$ln$T$ over a wide range, from 20 K to 1 K. 
 Previous studies reported a  $\chi \propto -\sqrt{T}$ behavior  below 4-1.4 K, suggesting a possible quadrupolar Kondo effect \cite{McElfresh_PRB_1993}.
  At 1 T, magnetic susceptibility $\chi(T)$ exhibits  a broad peak at approximately 1.3 K. However, this behavior is suppressed at higher fields, such as 8 and 13 T, where magnetic susceptibility becomes nearly constant, indicating a crossover from a NFL state to a Fermi-liquid (FL) state at high fields.
 
\color{black}

\begin{centering}
\begin{figure}[t]
\includegraphics[width=8.3cm]{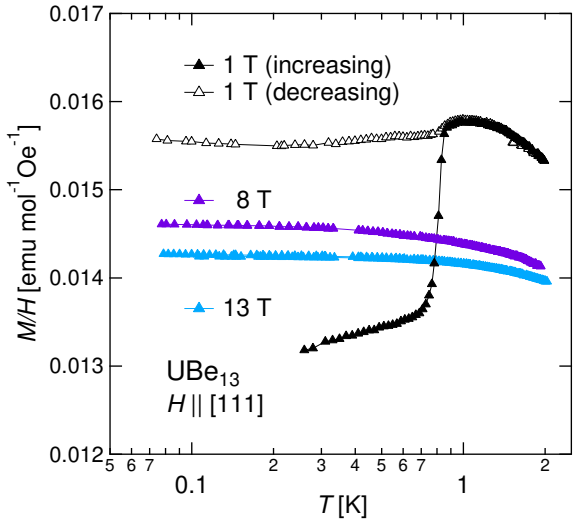}
\caption{   
Temperature dependence of  magnetic susceptibility ($\chi = M/H$) of UBe$_{13}$ along  $H ||$ [111] at 1, 8, and 13 T.
 }
\end{figure}
\end{centering}

\begin{centering}
\begin{figure}[t]
\includegraphics[width=8.8cm]{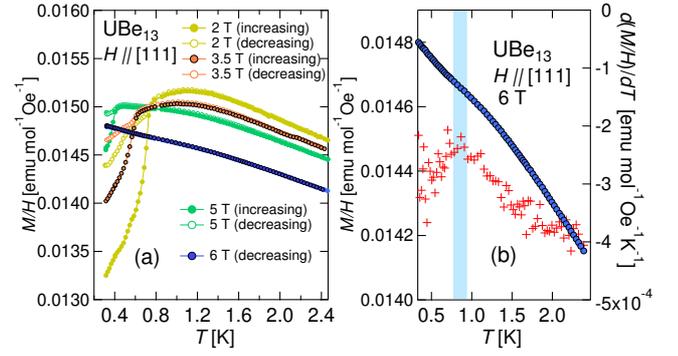}
\caption{   
(a) Temperature dependence of magnetic susceptibility ($\chi = M/H$) of UBe$_{13}$ along $H ||$ [111]  
 at 2, 3.5, 5, and 6 T.
(b) Temperature dependence of $\chi = M/H$ at 6 T ($H ||$ [111]) with its temperature derivative $d(M/H)/dT$.
 Here, the light-blue shaded region denotes the observed  anomaly  ($T_{\rm A}^{\chi }$) in the normal state,  
  defined  as the maximum in $d(M/H)/dT$.
 }
\end{figure}
\end{centering}

  A similar crossover to an FL behavior with increasing field  is observed along $H ||$ [111] (Fig. 4). At 1 T, $\chi(T)$ reaches its maximum at approximately 1.3 K. We define this characteristic temperature as $T_{\chi \rm max}$.

\color{black}
 
\begin{centering}
\begin{figure}[t]
\includegraphics[width=8cm]{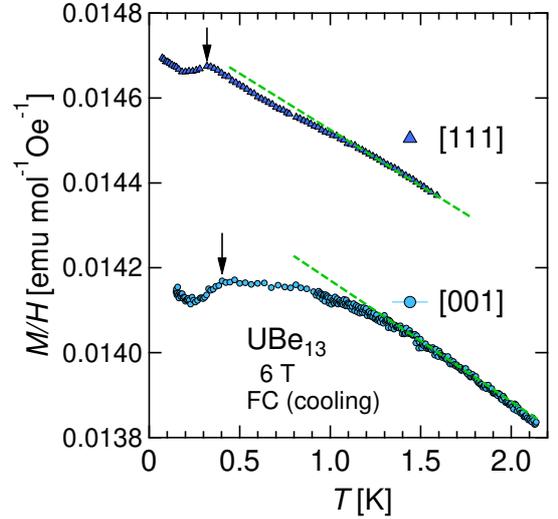}
\caption{   
Temperature dependence of $\chi = M/H$ at 6 T  applied   along $H ||$ [001] and [111]  (FC process), with arrows indicating superconducting transition at 6 T.
 }
\end{figure}
\end{centering}

Figure 5(a) presents the temperature dependence of the magnetic susceptibility of UBe$_{13}$ in the intermediate-field region (2 $\leq \mu_{0} H \leq$ 6 T) in $H ||$ [111]. At low temperatures, both the ZFC and FC processes exhibit kinks associated with the superconducting transition at 2, 3.5, and 5 T. In the normal state, a $\chi_{\rm max} $ anomaly is observed at 2 and 3.5 T. $T_{\chi \rm max}$ shifts slightly to lower temperatures from 2 T to 3.5 T but disappears at 5 T. By contrast, the $\chi_{\rm max} $ anomaly in $H ||$ [001] is rather insensitive to magnetic field, slightly shifting to higher temperatures (Supplmentary Materials, Fig. S1) 
 \cite{SupplementaryMaterials}. 
  The temperature dependence of $\chi = M/H$ exhibits a particularly interesting feature at 6 T: its temperature derivative ($dM/dT$, plotted on the right axis in Fig. 5(b)) shows a pronounced maximum at  $T_{\rm A}^{\chi} \simeq $ 0.8 K. This corresponds to an upturn in $\chi(T)$ at low temperatures (left axis in Fig. 5(b)). However, $\chi(T)$ exhibits nearly $T$-independent behavior at 8 T (Fig. 4). Such behavior cannot be explained by nuclear magnetization; instead, it suggests that the $5f$-electron system undergoes a nontrivial magnetic response in the intermediate-field region of 4-6 T.

 Here, we comment on the anisotropy of the anomalous upturn at 6 T. Figure 6 presents magnetic susceptibility ($M/H$) along $H ||$ [001] and $H ||$ [111] under FC conditions, where the downward arrows indicate superconducting transitions and the broken lines are guides. Interestingly, the low-temperature upturn of $\chi(T)$ is absent along $H ||$ [001]. For this field orientation, susceptibility shows near-saturation behavior just above the superconducting transition at 6 T. 
Indeed, magnetic susceptibility exhibits pronounced anisotropy between the two directions at 6 T.
\color{black}

 The above observations raise the question of how such anisotropy is reflected in the $M(H)$ curves. Figure 7 shows the magnetization curves at 0.13 K along both the [001] and [111] directions. The upper critical field $H_{\rm c2}$ is slightly higher along [001] than along [111]. For reference, the magnetization curve at $T =$ 0.82 K along $H ||$ [111] is also plotted. In UBe$_{13}$, the normal-state magnetization along [111] is larger than that along [001] at low temperatures. Here, the value of magnetization at 14.5 T is consistent with a previously reported high-field magnetization curve up to 60 T \cite{Detwiler_PRB_2000}. At first glance, the $M(H)$ curve in the normal state appears featureless in UBe$_{13}$. However, a detailed analysis of the magnetization curves reveals the presence of nonlinear contributions.

\begin{centering}
\begin{figure}[t]
\includegraphics[width=8.8cm]{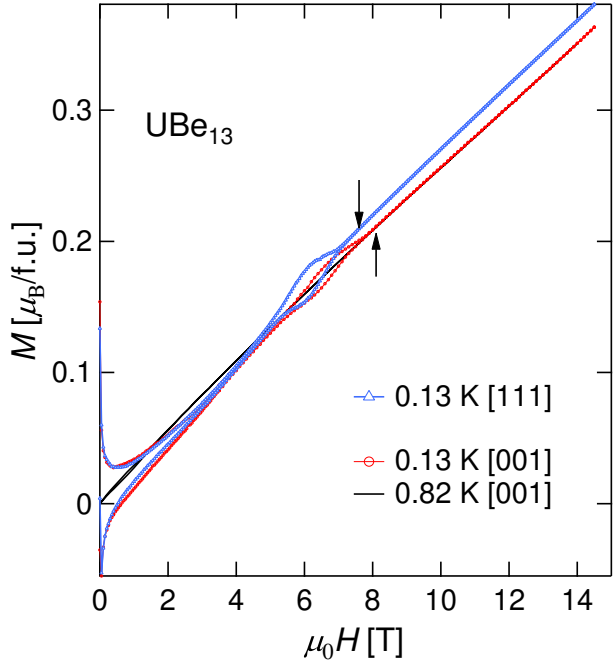}
\caption{   
Magnetization curves of UBe$_{13}$ single crystal at 0.13 K along $H ||$ [001] and $H ||$ [111], with arrows denoting  $H_{\rm c2}$.
 }
\end{figure}
\end{centering}

 In general, the magnetization of a paramagnet can be expressed as $M= \chi_{1} H + \chi_{3} H^3+ \chi_{5} H^5$ + $\ldots$. Accordingly, $M/H = \chi_{1} + \chi_{3} H^2+ \chi_{5} H^4$ + $\ldots$. Thus, the slope of $M/H$ when it is plotted as a function of $H^{2}$ corresponds to $\chi_{3}$. If the $M/H$ curve deviates from linearity against $H^{2}$, then higher-order contributions will be present, such as $\chi_{5}$. Figure 8 shows $M/H$ versus $H^{2}$ for UBe$_{13}$ along the [001] and [111] directions, as obtained from the equilibrium magnetization curves. At  0.8 K, the slopes of the $M/H$--$H^{2}$ plots show no anisotropy below 4 T, consistent with the findings of Ramirez $et$ $al.$ \cite{Ramirez_PRL_1994}.
 \color{black}

\begin{centering}
\begin{figure}[t]
\includegraphics[width=8.6cm]{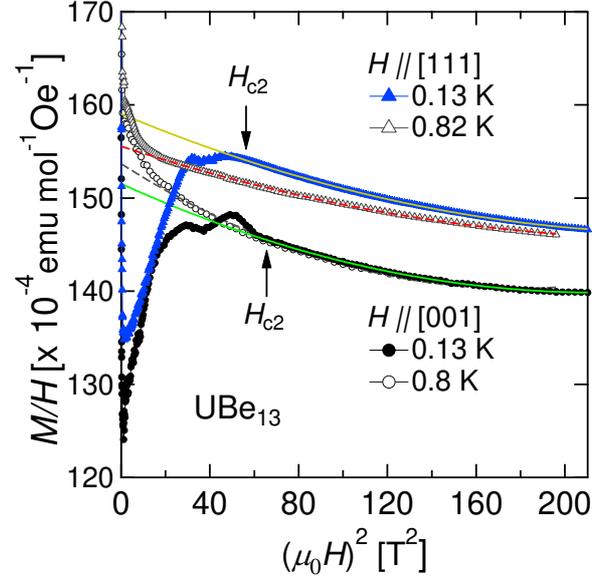}
\caption{   
$M/H$ vs. $H^{2}$ plot for equilibrium magnetization in UBe$_{13}$ at 0.13 and 0.8 K along $H ||$ [001] and $H ||$ [111], with arrows denoting upper critical field $H_{\rm c2}$.
 The solid and dashed lines represent the fitting results for 0.13 K and 0.8 K, respectively.
 }
\end{figure}
\end{centering}

\begin{centering}
\begin{figure}[t]
\includegraphics[width=7cm]{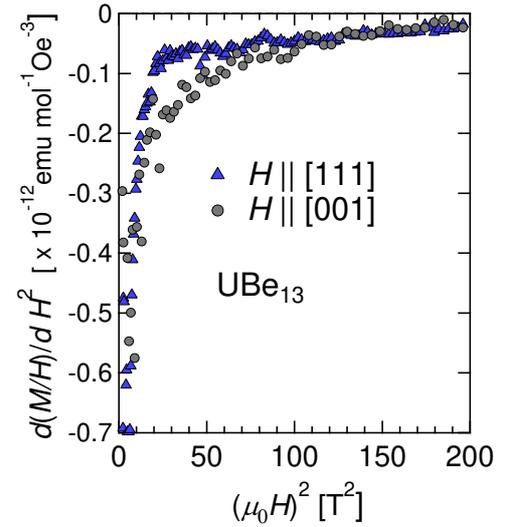}
\caption{   
Derivative of $M/H$ with respect to $H^2$ for UBe$_{13}$ single crystal  along $H ||$ [001] ($T = $ 0.8 K) and $H ||$ [111] ($T = $ 0.82 K).
 }
\end{figure}
\end{centering}

Interestingly, at above 4 T (Fig. 8), the slopes of the $M/H$--$H^{2}$ plots exhibit anisotropy and change with  increasing field. Thus, at above 4 T, $\chi_{3}$ becomes anisotropic and higher-order contributions ($\chi_{5}$) are nonnegligible. At 0.13 K, the upper critical field is approximately 8 T (as indicated by the arrows), where the $M/H$ curves show kinks along both field directions. However, in the normal state above 8 T, nonlinear components similar to those at 0.8 K are present in both directions.
 Using the relation  $M/H = \chi_{1} + \chi_{3}H^{2} + \chi_{5}H^{4}$, the $M(H)$ curves at 0.8 K can be well fitted above 4 T, and those at 0.13 K can also be fitted over a wide field range in the normal state above $H_{\rm c2}$.
  The solid and dashed lines in Fig. 8  represent the fitting results for 0.13 K and 0.8 K, respectively.
 For $H ||$ [001], the normal-state magnetization curves (above $H_{c2}$) behave almost identically at 0.8 and 0.13 K; along $H ||$ [111], the nonlinear components above $H_{\rm c2}$ exhibit a clear  temperature dependence.
 \color{black}

The nonlinear susceptibility $\chi_{5}$ is obtained from the slope of the derivative of $M/H$ as a function of $H^{2}$, as $\frac{d(M/H)}{d (H^2)} = \chi_{3} + 2 \chi_{5} H^2$ + $\ldots$. Figure 9 shows the results at 0.8 K. $\chi_{5}$ behaves anisotropically at 4--10 T. Below 5 T, $\chi_{5}$ is larger along $H ||$ [111] than along $H ||$ [001]. Above 5 T, $\chi_{5}$ along $H ||$ [111] becomes nearly constant ($\chi_{5} >$ 0). By contrast, along $H ||$ [001], $\chi_{5}$ changes gradually with increasing field  and approaches the value along $H ||$ [111] above 10 T.

\subsection{Specific-heat results  for  UBe$_{13}$} 

Specific-heat measurements were conducted to confirm whether the anomalies and anisotropy observed in the above  magnetization resluts are intrinsic. Figure 10 shows the temperature dependence of $C/T$ at zero field and at 14 T applied along the [001] and [111] directions. At zero field, $C/T$ does not saturate at low temperatures but exhibits an NFL behavior. At 14 T, the normal-state $C/T$ is suppressed, and this behavior shows pronounced anisotropy. 
 The specific heat is larger along $H ||$ [001], whereas along $H ||$ [111], the NFL behavior in $C/T$ is remarkably reduced, approaching the FL regime    at 14 T.
\color{black}

\begin{centering}
\begin{figure}[t]
\includegraphics[width=8.7cm]{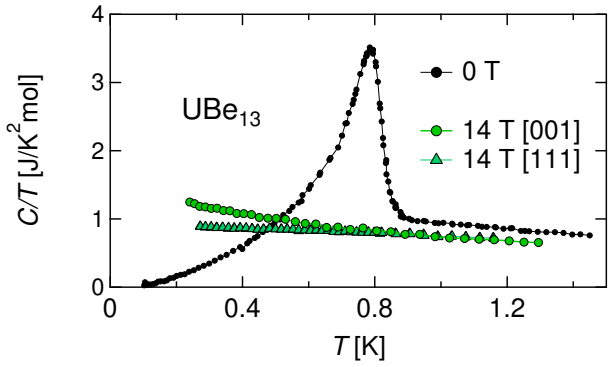}
\caption{   
Temperature dependence of the specific heat $C/T$ of UBe$_{13}$ at zero field and at 14 T applied along [001] and [111] directions.
 }
\end{figure}
\end{centering}

The nuclear contribution from $^{9}$Be nuclei at low temperatures has to be considered. The nuclear-spin relaxation rate in UBe$_{13}$ is 
 very small at low temperatures ($1/T_{1} \sim 10^{-3}$ s$^{-1}$---i.e., very long $T_{1}$) \cite{MacLaughlin_PRL_1984, Tien_PRB_1989}, suggesting that the nuclear contributions may be ineffective within our measurement timescale. In our quasi-adiabatic method, the typical measurement time is on the order of several hundred seconds. Calculations show that the nuclear specific heat of Be becomes nonnegligible below approximately 0.2 K relative to the electronic specific-heat value of UBe$_{13}$. Below 0.2 K, whether the observed temperature dependence originates purely from the $5f$-electron system is very difficult to determine (an evaluation of the nuclear specific heat is provided in  Supplementary Material) 
 \cite{SupplementaryMaterials}.  
  However, when the nuclear contribution is still small at above 0.24 K, the electronic specific heat can be quantitatively obtained, even at high fields. At 0.24 K, the value of $C/T$ along $H ||$ [111] is approximately 20$\%$ smaller than that along $H ||$ [001], demonstrating that the change in the density of states associated with the NFL-to-FL crossover is more remarkable along $H ||$ [111].
\color{black}

\begin{centering}
\begin{figure}[t]
\includegraphics[width=8.5cm]{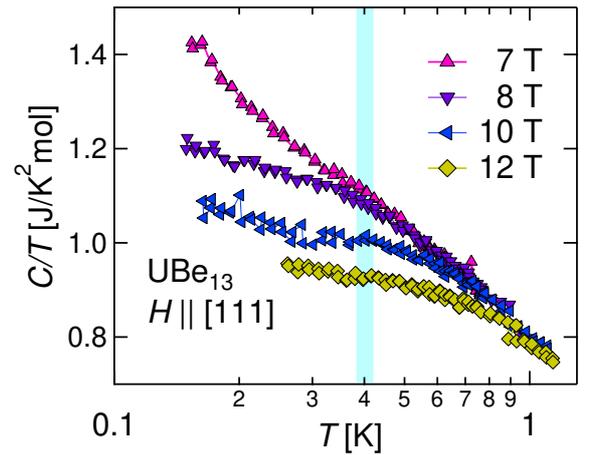}
\caption{ 
Temperature dependence of specific heat of UBe$_{13}$ along $H ||$ [111] at 7, 8, 10, and 12 T as function of the temperature on a logarithmic scale. The light-blue-shaded region denotes the onset of the crossover anomaly ($T_{\rm A}^{[111]}$) in the normal state, appearing at approximately 7 T.
}
\end{figure}
\end{centering}

A striking feature is the considerable anisotropy in specific heat at high fields (Fig. 10). 
Such pronounced anisotropy in specific heat has not been reported in correlated-electron systems, and 
  it is anomalous behavior unique to UBe$_{13}$. For example, in the tetragonal heavy-electron superconductor CeCoIn$_{5}$, no comparable anisotropy has been reported in the normal state near the superconducting state \cite{Sakakibara_RPP_2016}. Similarly, in the hexagonal heavy-electron superconductor UPd$_{2}$Al$_{3}$, no anisotropy in specific heat has been observed within the $ac$ plane \cite{Shimizu_PRL_2016} above the upper critical field.
\color{black}

Figure 11 shows the $C(T)/T$ measured at 7--12 T along $H ||$ [111]. At 7 and 8 T, $C/T$ shows a slight kink at  0.4--0.5 K. This feature is likely associated with the anomaly ($T_{\rm A}^{\chi}$) observed in the temperature dependence of magnetic susceptibility along $H ||$ [111] (Figs. 5 and 6). With increasing field, $C/T$ is suppressed, tending to become nearly temperature independent at 12 T. Therefore, along $H ||$ [111], a thermodynamic anomaly, $T_{\rm A}$, emerges in the normal state   through the NFL-to-FL crossover.
\color{black}

Next, we present the field dependence of specific heat. Figure 12 shows $C(H)/T$ along $H ||$ [111] at 0.24, 0.41, 0.60, 0.80, 0.94, and 1.20 K. Between 0.24 and 0.80 K, a clear jump in specific heat is observed below 7 T, corresponding to $H_{\rm c2}$; the magnitude of this jump decreases with temperature. The inset displays an expanded view of the normal-state region of these data. Importantly, the normal-state behavior above $\sim$ 6 T shows distinctive features, with $C(H)/T$ exhibiting a broad maximum at 0.60 K in 7 T. We define this anomaly as $H_{\rm A}$. At 0.94 and 1.20 K, $C(H)/T$ decreases with increasing field  below 5 T, 
  and the high-field anomaly ($H_{\rm A}$)  is observed  at around 9 T. 
 As the temperature decreases, this anomaly shifts to lower fields. At 0.40 K, it shifts further and approaches $H_{\rm c2}$ closely.
\color{black}

\begin{centering}
\begin{figure}[t]
\includegraphics[width=9cm]{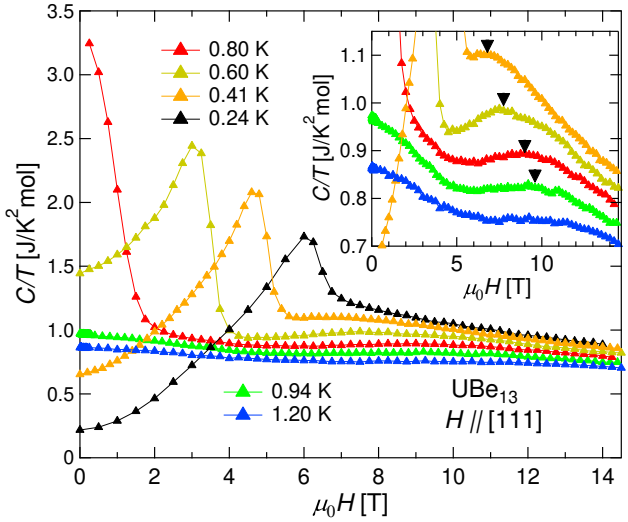}
\caption{ 
Field dependence of specific heat $C/T$ of UBe$_{13}$ along $H ||$ [111] at $T = $ 0.24, 0.41, 0.60, 0.80, 0.94, and 1.20 K. Inset: expanded view of normal-state data, with arrows denoting high-field anomaly ($H_{\rm A}$) at approximately 7--10 T.
}
\end{figure}
\end{centering}

\begin{centering}
\begin{figure}[t]
\includegraphics[width=8.9cm]{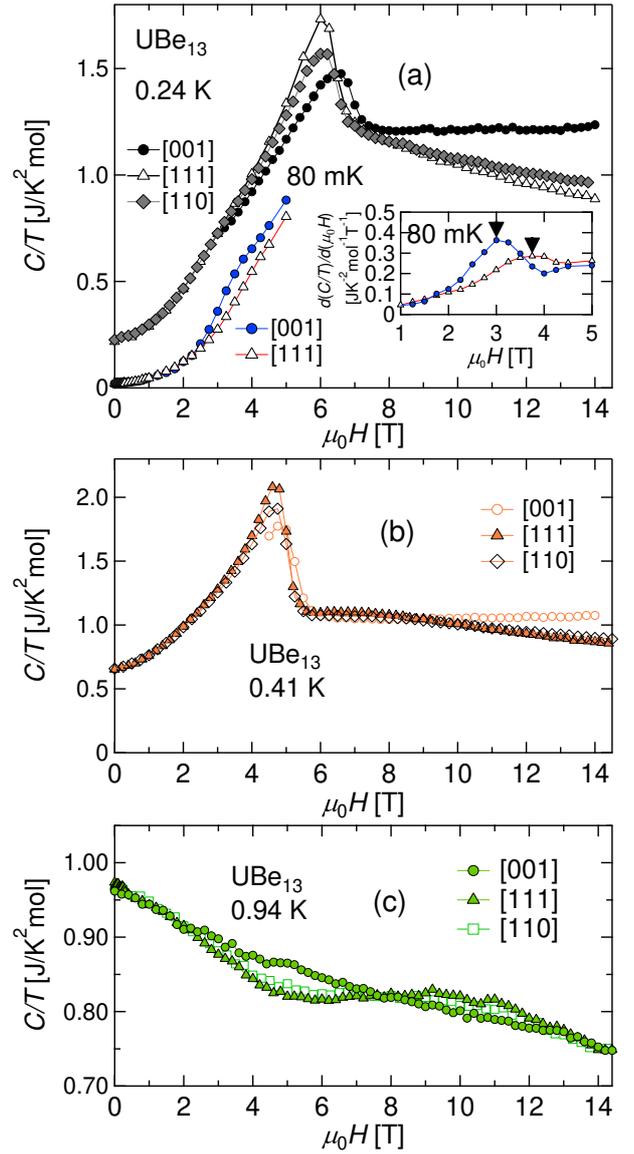}
\caption{   
Field dependence of specific heat $C/T$ of UBe$_{13}$ along $H ||$ [001], [111], and [110] at (a) $T = $ 0.08 and 0.24 K, (b) 0.41 K, and (c) 0.94 K. Inset: derivative of $C/T$ as function of $H$ at 0.08 K, with arrows indicating $B^{*}$ anomaly.
 }
\end{figure}
\end{centering}

In Figs. 13(a)--13(c), we compare the results obtained at fixed temperatures of 0.24, 0.41, and 0.94 K along [001], [111], and [110]. We also plot the $C(H)/T$ of UBe$_{13}$ at the lowest temperature (80 mK) along $H ||$ [001] and [111] in Fig. 13(a) \cite{Shimizu_JMMM_2016}. Here, due to the difficulty of analyzing nuclear specific heat at 0.08 K, we show reliable data up to 5 T at this temperature. For the superconducting state, we observe the $B^*$ anomaly 
 at 3--4 T \color{black} at 0.08 K along $H ||$ [001] and [111].
 We clarify the $B^*$ anomaly by plotting the $C/T$ derivative as a function of field  [inset of Fig. 13(a)]. In this paper, we define $B^*$ as a field where $d(C/T)/dH$ shows a peak (downward arrows). Interestingly, at 0.08 K, $C(H)/T$ exhibits strong anisotropy near the $B^*$ anomaly ($C_{[111]} < C_{[001] }$) [inset of Fig. 13(a)], but this anomaly becomes less pronounced at 0.41 K [Fig. 13(b)].

High-field anomalies in $C(H)/T$ are also detected above 7 T along $H ||$ [110]. Detailed $C/T$ data along $H ||$ [110] at various temperatures are provided in Supplementary Materials (Fig. S4) 
 \cite{SupplementaryMaterials}. By contrast, no such anomaly is observed along $H ||$ [001] [Figs. 13(a)--13(c)]. At 0.94 K, the high-field anomaly is seen in the normal-state specific heat along $H ||$ [111], whereas $C(H)/T$ decreases monotonically along $H ||$ [001] [Fig. 13(c)]. Despite the small anisotropy of the upper critical field $H_{\rm c2}$ in UBe$_{13}$, we observe its anisotropy: $H_{\rm c2}^{[110]} < H_{\rm c2}^{[111]} < H_{\rm c2}^{[001]}$. In this situation, a similar relation in specific-heat peaks along the three field directions is expected. However, the $C(H)/T $ peak is largest along $H ||$ [111]. As the high-field anomaly at $H_{\rm A}$ is most pronounced along $H || $[111], the superconducting state may be enhanced in association with changes in the electronic state at the applied fields along [111].

\subsection{$H$--$T$ phase diagram of  UBe$_{13}$} 

Figure 14 shows the $H$--$T$ phase diagram of the UBe$_{13}$ single crystal obtained from the present magnetization and specific-heat measurements. 
 Here, $H_{\rm c2}$ and $T_{\rm sc}$ are defined  as the midpoints of the step-like (or peak) anomalies in 
  $C(H)/T$ and $C(T)/T$ curves, respectively. This definition is consistent with the results obtained from the magnetization measurements. 
Regarding $H_{\rm c2}$, no anisotropy is observed at low fields, but anisotropy develops above 5 T. This behavior is consistent with previous reports \cite{Shimizu_JPSJ_2011, Shimizu_PRB_2016}. Among the field orientations, $H_{\rm c2}$ is largest along [001]. 
 Such anisotropy may originate from the superconducting gap symmetry or the anisotropy of the Fermi surface.
\color{black}

Regarding the normal state in the $H$--$T$ phase diagram, the susceptibility maximum temperature ($T_{\chi \rm max}$) shifts slightly to higher temperatures and then moves toward lower temperatures along $H ||$ [111] with increasing  field. Along $H ||$ [001], $T_{\chi \rm max}$ slightly shifts to higher temperatures with increasing field. This susceptibility maximum corresponds to the so-called 2 K anomaly previously observed in resistivity \cite{Ott_PRL_1983, Schmiedeshoff_PRB_1996}, specific heat \cite{Kim_PRB_1995, Kromer_PRB_2000}, thermal expansion \cite{Kromer_PRB_2000}, and thermoelectric power \cite{Shimizu_PRBR_2015}. The 2 K anomaly is likely associated with the freezing of $5f$-electron degrees of freedom by the Kondo effect, or the onset of its coherence behavior. Interestingly, along $H ||$ [111], $T_{\chi \rm max}(H)$ appears to meet the high-field anomaly $T_{\rm A}(H)$ in the phase diagram. 
 By contrast, along $H ||$ [001], no high-field anomaly ($T_{\rm A}^{\chi}$ and $H_{\rm A}$) is observed, implying  a close  relationship between the suppression of the susceptibility maximum ($T_{\chi \rm max}$) toward lower temperatures  and the emergence of the high-field anomaly $T_{\rm A}^{\chi}$.
\color{black}

\begin{centering}
\begin{figure}[t]
\includegraphics[width=9cm]{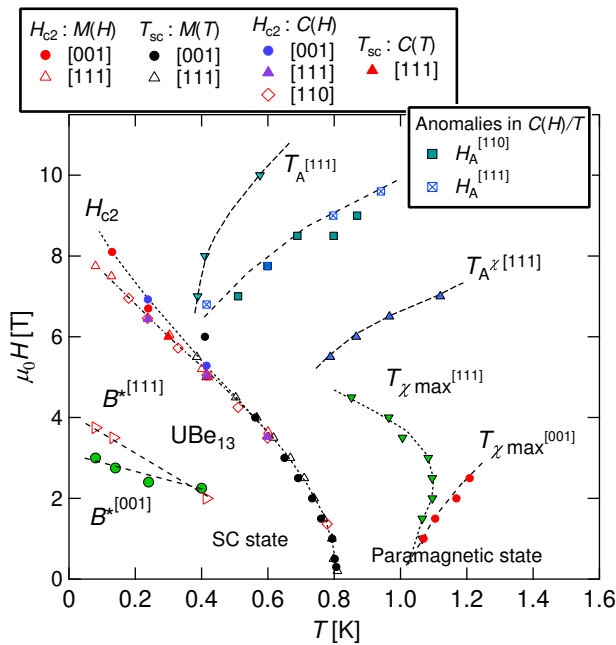}
\caption{  
  $H-T$ phase diagram of UBe$_{13}$ single crystal. The diagram includes the superconducting upper critical field, $B^{*}$ anomaly, and thermodynamic anomalies observed in the normal state. Here, $T_{\chi \rm max}$ and $T_{\rm A}$ denote the temperatures of the susceptibility maximum and the high-field anomaly above $\sim$ 6 T, respectively. In addition, $H_{\rm A}$ is defined by the field dependence of the specific heat, $C(H)/T$.
 }
\end{figure}
\end{centering}

\section{Discussion}
\subsection{Superconducting  state in UBe$_{13}$} 

We first discuss the superconducting state of UBe$_{13}$. The $H_{c2}(0)$ of UBe$_{13}$ is much larger than the Pauli-limiting field, and experimental results supporting odd-parity pairing have been published \cite{Ott_PRL_1984, MacLaughlin_PRL_1984, Einzel_PRL_1986, Tien_PRB_1989}. Nevertheless, $H_{\rm c2}$ shows strong bending at approximately 3 T
 and reduction of  the Ginzburg-Landau parameter $\kappa_{2}(T)$ upon cooling, indicating the presence of a Pauli limit in UBe$_{13}$ \cite{Shimizu_JPSJ_2011}. Regarding the superconducting gap structure, the isotropic linear-$H$-like behavior in $C(H)/T$ at the lowest temperature [Fig. 13(a)] strongly indicates a fully gapped superconducting state in UBe$_{13}$ \cite{Shimizu_PRL_2015}.
\color{black}

In the case of odd-parity superconducting states in a cubic symmetry with strong spin--orbit coupling, fully gapped states are possible for $A_{1u}$ (one dimensional), $E_u$ (two dimensional), and $T_{2u}$ (three dimensional) symmetries \cite{Blount_PRB_1985}. Experiments on a polycrystalline sample showed that $H_{\rm c2}$ can be well reproduced by the $A_{1u}$ state at ambient pressure \cite{Shimizu_PRL_2019}. In the $A_{1u}$ state, the Pauli effect isotropically occurs. The presence of the isotropic Pauli effect below 3 T has been indicated by results of specific-heat and magnetization measurements \cite{Shimizu_JPSJ_2011, Shimizu_PRB_2016}. By contrast, at high fields, the spin component will be parallel to the field direction and the Pauli effect will be absent. Thus, the paramagnetic limiting below 3 T and the absence of a Pauli limit at higher fields can be explained within the framework of the $A_{1u}$ state \cite{Fomin_JLTP_2000, Shimizu_PRL_2019}.
\color{black}

For one of the $E_{u}$ states, $k_{x} \bm{ \hat{x}} + k_{y} \bm{ \hat{y} } - 2 k_{z} \bm {\hat{z} }$, the superconducting gap is fully open and the magnitude of the $\bm{d}$-vector is finite along all directions \cite{Blount_PRB_1985, Machida_JPSJ_2018}. Therefore, the paramagnetic effect is also expected along any field direction in this nematic-type $E_{u}$ \cite{Machida_JPSJ_2018} state; this state is also a plausible explanation for the superconducting gap symmetry in UBe$_{13}$ and Th-doped system \cite{Shimizu_PRBR_2017, Machida_JPSJ_2018}.
\color{black}

 UBe$_{13}$ exhibits the $B^{*}$ anomaly in the superconducting state only. Importantly, this anomaly has been observed in both Al-flux-grown single crystals and arc-melted polycrystalline samples \cite{Ellman_PRBR_1991, Kromer_PRL_1998, Langhammer_JMMM_1998, Walti_PRBR_2001, Kromer_PRB_2000, Shimizu_PRL_2012, Shimizu_PRB_2016}. Because the $B^{*}$ anomaly appears as a thermodynamic anomaly in specific heat and magnetization, it is considered to originate either from a change in the superconducting state or from a variation in the electronic density of states inside vortex cores. For $A_{1u}$ and nematic-type $E_{u}$, where the Pauli effect is present at low fields along all directions, the alignment of the Cooper-pair spins along the magnetic field can induce the $B^{*}$ anomaly. Here, the quasiparticle excitations of $C(H)/T$ is weaker along $H$ $||$ [111] near $B^{*}$ than along $H$ $||$ [001] [Fig. 13(a)]. This phenomenon may be explained by the anisotropic Fermi surface: the Fermi surface is  absent along the $\langle 111 \rangle$ direction except for a tiny electron pocket, according to previous band calculations \cite{Takegahara_Physica_2000, Maehira_PhysicaB_2002}.
\color{black}


\subsection{Unusual normal state in UBe$_{13}$} 

 We have uncovered  multiple thermodynamic anomalies in the normal state of UBe$_{13}$ near the superconducting phase: $T_{\chi \rm max}$, $T_{\rm A}$, and  $H_{\rm A}$ \cite{Endnote1} (Fig. 14). High-field anomalies were also observed in magnetic-torque measurements of polycrystalline UBe$_{13}$, with a similar temperature dependence \cite{Schmiedeshoff_PRB_1993}. In addition, previous thermoelectric power measurements on polycrystalline UBe$_{13}$ revealed high-field anomalies at 7--12 T \cite{Shimizu_PRBR_2015}, suggesting Fermi-surface reconstruction above 7 T. The clear detection of these high-field anomalies along [111] and [110] but not along [001] indicates anisotropic Fermi-surface reconstruction at high fields [Figs. 13(a)-(c)]. If, as recently proposed, the NFL behavior in UBe$_{13}$ originates from a two-channel Kondo effect associated with the semimetallic character \cite{Iimura_PRB_2019},  this NFL state will be strongly affected by such Fermi-surface reconstruction.
 
 \color{black}

Our  results indicate that the above anisotropic high-field anomaly ($T_{\rm A}$ and $H_{\rm A}$) is closely related to the observed nonlinear susceptibility. Detailed studies on nonlinear susceptibilities up to the fifth order have  been reported for the heavy-fermion superconductor UPt$_{3}$ \cite{Shivaram_PRB_2014, Shivaram_PRB_2014_5th}, which shows sign reversals of $\chi_{3}(T)$ and $\chi_{5}(T)$ near the temperature where an itinerant metamagnetic transition occurs. By contrast, in UBe$_{13}$, no such $\chi_{3}$ and $\chi_{5}$ sign reversal is observed with temperature ($\chi_3 < $ 0, $\chi_5 > $ 0). Thus, the origin of the nonlinear susceptibilities in UBe$_{13}$ is distinct from that associated with itinerant metamagnetism in UPt$_{3}$.
 \color{black}

The nonlinear susceptibility $\chi_{3}$ of UBe$_{13}$ was precisely investigated below 4 T in an earlier study \cite{Ramirez_PRL_1994}; they  were motivated to test the possibility of a quadrupolar Kondo effect \cite{Cox_PRL_1987, Cox_JPhysC_1996}, which occurs for the non-Kramers doublet $\Gamma_{3}$ ($5f^2$, $J =$ 4) in the CEF ground state. However, the $\chi_{3}(T)$ results for UBe$_{13}$ do not support the non-Kramers doublet in the CEF ground state \cite{Ramirez_PRL_1994}. 
  Nevertheless,  \color{black} 
   the possible occurrence of the quadrupole Kondo effect in U$_{0.9}$Th$_{0.1}$Be$_{13}$ was proposed  from nonlinear ($\chi_{3}$) susceptibility measurements \cite{Aliev_EPL_1995, Aliev_JPhysC_1996}.
 Therefore, the nature of UBe$_{13}$  has not yet been fully established and remains controversial.

According to another theoretical proposal, if the CEF ground state of UBe$_{13}$ is the $\Gamma_{1}$ singlet ($5f^2$, $J =$ 4), then the competition between the Kondo--Yosida singlet and the CEF singlet can  induce an NFL behavior resembling a two-channel Kondo effect; it may  also lead to magnetically robust specific heat \cite{Nishiyama_JPSJ_2011}. If $\Gamma_{1}$ is the CEF ground state in UBe$_{13}$, then the first excited state will be either $\Gamma_{5}$ (triplet) or $\Gamma_{4}$ (triplet) \cite{Lea_JPhysChemSolids_1962}. With $c$-$5f$ hybridization effects, these excited CEF levels can couple to conduction electrons \cite{Nishiyama_JPSJ_2011}. Here, the $\Gamma_{5}$ triplet possesses quadrupolar ($O_{xy}$, $O_{yz}$, $O_{zx}$) and magnetic octupolar  degrees of freedom. The $\Gamma_{4}$ triplet also has dipolar and magnetic octupolar degrees of freedom. In applied fields, these triplet states undergo Zeeman splitting; then, the excited states can influence the $5f$ electron ground-state properties.
 \color{black}

This study has elucidated the anomalous behavior of $\chi_{3}$ and $\chi_{5}$ in UBe$_{13}$ above 4 T. In particular, $\chi_{5}$ varies  
 clearly with temperature along $H ||$ [111]. A recent  theoretical study  identified magnetic octupolar effects in fifth-order  nonlinear susceptibility  $\chi_{5}$ \cite{Sorensen_PRB_2021}. The anisotropy in $\chi_{5}$ observed in UBe$_{13}$ in the current work may thus be interpreted as a field-induced higher-order multipolar effect under $O_{h}$ symmetry. Furthermore, the observed anomalies along $H ||$ [111] and [110] indicate that the reconstruction of the $5f$-electron density of states (Fermi surface) may be strongly influenced by the field-induced multipolar correlations in UBe$_{13}$. 
 \color{black}

\section{Summary}

The anomalous superconducting and NFL states of UBe$_{13}$ were studied using high-resolution dc magnetization and heat-capacity measurements down to 80 mK. Magnetization and heat-capacity data revealed the presence of a field-induced anomaly in magnetic fields  above 6 T along $H$ $||$ $[111]$. In the low-field region (below 4 T), we found a susceptibility maximum anomaly, which is associated with the freezing of $5f$ degrees of freedom in UBe$_{13}$. The field-induced anomaly observed above 6 T is remarkably anisotropic, suggesting the occurrence of anisotropic Fermi-surface reconstruction in UBe$_{13}$. Moreover, anomalous nonlinear fifth-order susceptibility was found near the superconducting state in UBe$_{13}$, implying a possible relationship between high-rank multipolar degrees of freedom and the low-temperature magnetic response. In the superconducting state, we observed a remarkable anisotropy around the $B^{*}$ anomaly, namely, $C_{[111]} < C_{[001]}$, which is the opposite of the anisotropy observed in the normal state near $T_{\rm A}$ and $H_{\rm A}$. These observations regarding anisotropy in UBe$_{13}$ provide important insights into the interplay between the anomalous superconducting and normal states of this compound.

\begin{acknowledgments}
We are grateful to S. Hoshino and E. Svanidze  for valuable discussions. 
The present study was supported by
Graints-in-Aid KAKENHI (No. JP20K03851, JP22KK0224, JP23K03314, JP23H04870, JP23H04868, JP23K25825, JP21K03455, JP23H04871, JP23K03332, JP23K25829) from the Ministry of Education, Culture, Sports, Science and Technology (MEXT) of Japan.
\end{acknowledgments}


\begin{thebibliography}{99}
\bibitem{Endnote1}  The anomalies observed in the temperature dependence of the specific heat ($T_{\rm A}$), in the field dependence of the specific heat ($H_{\rm A}$), and in the susceptibility ($T_{\rm A}^{\chi}$) appear at slightly different points in the phase diagram; this discrepancy is likely due to the differences in the definitions of  each physical quantity.
\end{thebibliography}


\end{document}